\documentclass[12pt]{iopart}
\usepackage{amssymb}
\usepackage{amstext}
\usepackage{color}
\usepackage{graphicx}

\newcommand{\um}[1]{\;\mathrm{#1}}
\newcommand{\ket}[1]{\left\vert #1\,\right\rangle}

\newcommand{\bra}[1]{\left\langle #1\,\right\vert}

\newcommand{\eqref}[1]{(\ref{#1})}

\begin{document}
\title[Optimal transport efficiency in models of light-harvesting complexes]{Optimal efficiency of quantum transport in structured disordered systems with applications to light-harvesting complexes}

% %%%%Infos for APS journals
% \date{\today}
% \author{Giulio~G. \surname{Giusteri}}
% \affiliation{Dipartimento di Matematica e Fisica and Interdisciplinary Laboratories for Advanced Materials Physics, Universit\`a Cattolica del Sacro Cuore, via Musei 41, I-25121 Brescia, Italy \\ and Istituto Nazionale di Fisica Nucleare, Sezione di Pavia, via Bassi 6, I-27100,  Pavia, Italy}
% \affiliation{International Research Center on Mathematics \& Mechanics of Complex Systems, via XIX marzo 1 I-04012 Cisterna di Latina, Italy}
% \author{G.~Luca \surname{Celardo}}
% \affiliation{Dipartimento di Matematica e Fisica and Interdisciplinary Laboratories for Advanced Materials Physics, Universit\`a Cattolica del Sacro Cuore, via Musei 41, I-25121 Brescia, Italy \\ and Istituto Nazionale di Fisica Nucleare, Sezione di Pavia, via Bassi 6, I-27100,  Pavia, Italy}
% \author{Fausto \surname{Borgonovi}}
% \affiliation{Dipartimento di Matematica e Fisica and Interdisciplinary Laboratories for Advanced Materials Physics, Universit\`a Cattolica del Sacro Cuore, via Musei 41, I-25121 Brescia, Italy \\ and Istituto Nazionale di Fisica Nucleare, Sezione di Pavia, via Bassi 6, I-27100,  Pavia, Italy}

%%%%%Infos for IOP journals

\author{G G Giusteri$^{1,2}$, G L Celardo$^{1}$ and F Borgonovi$^{1}$}
\address{$^1$Dipartimento di Matematica e Fisica and Interdisciplinary Laboratories for Advanced Materials Physics, Universit\`a Cattolica del Sacro Cuore, via Musei 41, I-25121 Brescia, Italy 
\emph{and}
Istituto Nazionale di Fisica Nucleare, Sezione di Pavia, via Bassi 6, I-27100,  Pavia, Italy}
\address{$^2$International Research Center on Mathematics \& Mechanics of Complex Systems, via XIX marzo 1 I-04012 Cisterna di Latina, Italy}
\ead{giulio.giusteri@unicatt.it}

\begin{abstract}
Disordered quantum networks, as those describing 
 light-harvesting complexes, are often characterized 
by the presence of antenna structures where the light 
is captured and inner structures (reaction centers) where
the excitation is transferred. Antennae often display distinguished
coherent features: 
their eigenstates  can be separated, with respect to the transfer of 
excitation, in the two classes of superradiant and subradiant states. 
Both are important to optimize transfer efficiency.
In absence of disorder  superradiant states have an enhanced coupling strength to the
RC, while subradiant ones are basically decoupled from it. 
Disorder induces a coupling between subradiant and superradiant
states, thus creating an indirect coupling to the RC. 
We consider the problem of finding the maximal  excitation transfer
 efficiency as a function of the RC energy and the disorder strength, first 
in a paradigmatic three-level system and then in  
 a realistic model for the light-harvesting 
complex of purple bacteria. 
Specifically, we focus on the case in which the excitation is
initially on  a subradiant state, showing that the optimal disorder is of
the order of the superradiant coupling. We also determine the
optimal detuning between the initial state and the RC energy.
We show that the efficiency remains high around the optimal
detuning in a large energy window, proportional to the superradiant
coupling. This allows for the simultaneous  optimization of excitation
transfer from several initial
states with different optimal detuning.

\end{abstract}


%\submitto{\NJP}%only for IOP

\maketitle

\section{Introduction}

Photosynthetic bacteria utilize antenna complexes to capture photons
and convert the energy of the short-lived electronic excitation in a more stable form,
such as chemical bonds. After absorption, the energy is
transferred to a complex, called reaction center (RC), where 
it initiates electron transfer, resulting in a
membrane potential. 
%This energy migration process has been 
%described by F\"orster \cite{forster1, forster2}
%and it is called Fluorescence Resonance Energy Transfer.
This very efficient transfer occurs on a time-scale of few hundreds of picoseconds and on a length-scale of few nanometers, so that  
coherent quantum dynamics can enter the play, as recent experiments seem to
prove~\cite{effnat}. 
Quantum coherence can
enhance transport efficiency inducing Supertransfer and Superradiance in light-harvesting complexes~\cite{srfmo,srrc,others}.
On the other hand, quantum coherence can also be
detrimental to transport, as Anderson localization~\cite{Anderson} and the presence of trapping-free subspaces~\cite{invsub} show.

Superradiance~\cite{Zannals,rottertb,puebla}, as viewed in
the context of both optical
fluorescence \cite{mukameldeph,robin,vangrondelle} and quantum
transport in open systems \cite{kaplan,srfmo,alberto,prbdisorder,prbdephasing}, is not solely a many-body effect. 
Single-excitation superradiance
is a prominent example of genuinely quantum cooperative
effect~\cite{scully}, relevant in natural complexes,
which operate in the single excitation regime since
solar light is very dilute.

Natural complexes are subject  to a noisy environment with different correlation time-scales (if compared to the excitonic transport time): 
(i) short-time correlations, giving rise to dephasing
(homogeneous broadening, as considered e.g.\ 
in \cite{enaqt6, enaqt3, enaqt4, enaqt9})
and (ii) long-time correlations, producing on-site static disorder
 (inhomogeneous broadening, as considered e.g.\ in \cite{disorder,fassioli}).
The role of environmental noise is twofold: on one hand,
it can help transport since it destroys the detrimental coherent
effects, leading to noise-enhanced energy transfer,
i.e.\ the existence of a maximal efficiency at some intermediate
noise strength, as found in the last decade by various groups
\cite{enaqt1,enaqt2,enaqt3,enaqt4,invsub,enaqt5,enaqt6,enaqt7,enaqt8,plenioPTA,enaqt9,caoprl}. On the other hand, it can suppress the beneficial coherences leading to a
quenching of Supertransfer~\cite{prbdisorder}. It is thus essential to
consider this non-trivial interplay when dealing with realistic light-harvesting
complexes, as it will be shown below.

The typical structure of bacterial photosynthetic complexes displays a
RC placed at the center of the light-harvesting complex I (LHI), with the
chromophores arranged on a ring and surrounded by other ring
structures (called LHII) acting as peripheral antennae.
The LHI-RC structure describing light-harvesting complexes produces a distinguished feature:
the eigenstates of the peripheral ring structure can be separated into
two classes, superradiant and subradiant states, with respect to
the transfer of excitation towards the RC. In absence of disorder, the superradiant states are
coupled to the RC with a coupling amplitude proportional to
$\sqrt{N}$, where $N$ is the number of chromophores in the ring, while
subradiant states are basically decoupled from the RC. 
In presence of disorder, the subradiant states can be coupled to the
superradiant ones and, as a consequence, only indirectly coupled to
the RC states. Note that the subradiant subspace, previously studied
by the authors~\cite{prbdisorder}, has been also analyzed in the
literature under the name of trapping-free subspace~\cite{invsub}.

Here we want to discuss the optimization of excitation transfer
efficiency from peripheral states in networks displaying the Ring-RC
structure, with particular reference to subradiant initial states. 
Indeed, since disorder is only detrimental for the superradiant state,
we expect that the optimal condition will be mainly determined by the
behavior of subradiant states. 
Nevertheless, the role of superradiant states is very important \cite{prbdisorder} and it will be the subject of future investigations.

Even if in natural systems the role played by dephasing is crucial,
here we will focus only on the interplay of
static disorder and energy landscape (energy detuning between the
initial state and the RC) on transport efficiency.  

To pursue our goal, we consider in section~\ref{sec:trimer} a paradigmatic model:
a trimer displaying a superradiant and a subradiant state, on which the
excitation is initialized, and an acceptor state (the RC), where the
excitation can be trapped, see figure~\ref{trimer}. 
In absence of disorder, while the subradiant state is decoupled from the RC, the superradiant
state has an enhanced coupling $V_{\rm rc}$  to the RC. 
We discuss the maximization of the average (over disorder) transfer
efficiency with respect to both the detuning $\delta E_{\rm rc}$ between initial
subradiant state and
RC and the strength $W$ of disorder. 
We have found two different regimes depending on 
$V_{\rm rc}$ and the energy gap $\Delta$ between the subradiant and superradiant states. 
While the optimal noise is always of the order of $V_{\rm rc}$, 
the optimal detuning occurs at resonance with the initial energy, 
$\delta E_{\rm rc} = 0$,
only for $V_{\rm rc} > \Delta$. Another optimal detuning, 
$\delta E_{\rm rc} = V_{\rm rc}^2/\Delta$,
% intrinsically non-perturbative and based on the concept of dressed states
should be used in the case $V_{\rm rc} \lesssim \Delta$.

In section~\ref{sec:natcomp} we show the relevance of our findings in a more realistic model by analyzing the LHI-RC complex of purple bacteria as presented in~\cite{schultenH}.
After a reduction of the whole system to simpler systems capable to describe its main features, we prove that the optimal conditions for the excitation 
transfer from a room-temperature Gibbs 
distribution of ring states can be estimated by applying our analysis. 
The optimal disorder strength we find falls within the estimated range of natural disorder strength.
We conclude the paper (section~\ref{sec:conclusions}) with a summary of the main results and of future research directions.

\section{The trimer model}

\label{sec:trimer}
\begin{figure}
\begin{indented}\item[]
\includegraphics[width=5.5cm]{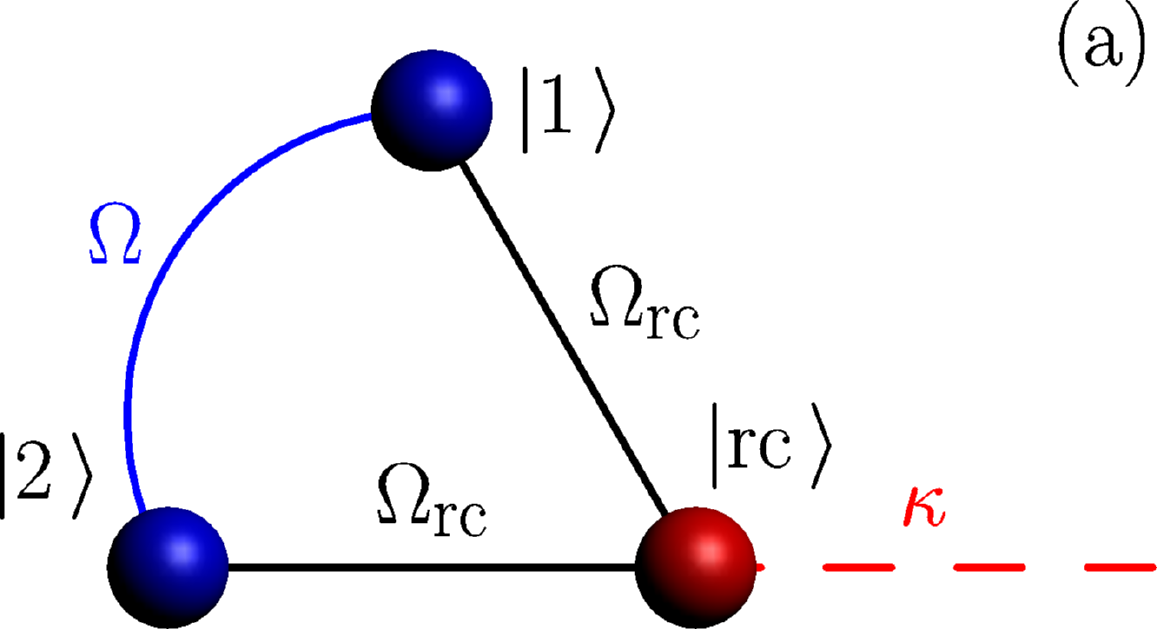}\hspace{1.cm}
\includegraphics[width=5.5cm]{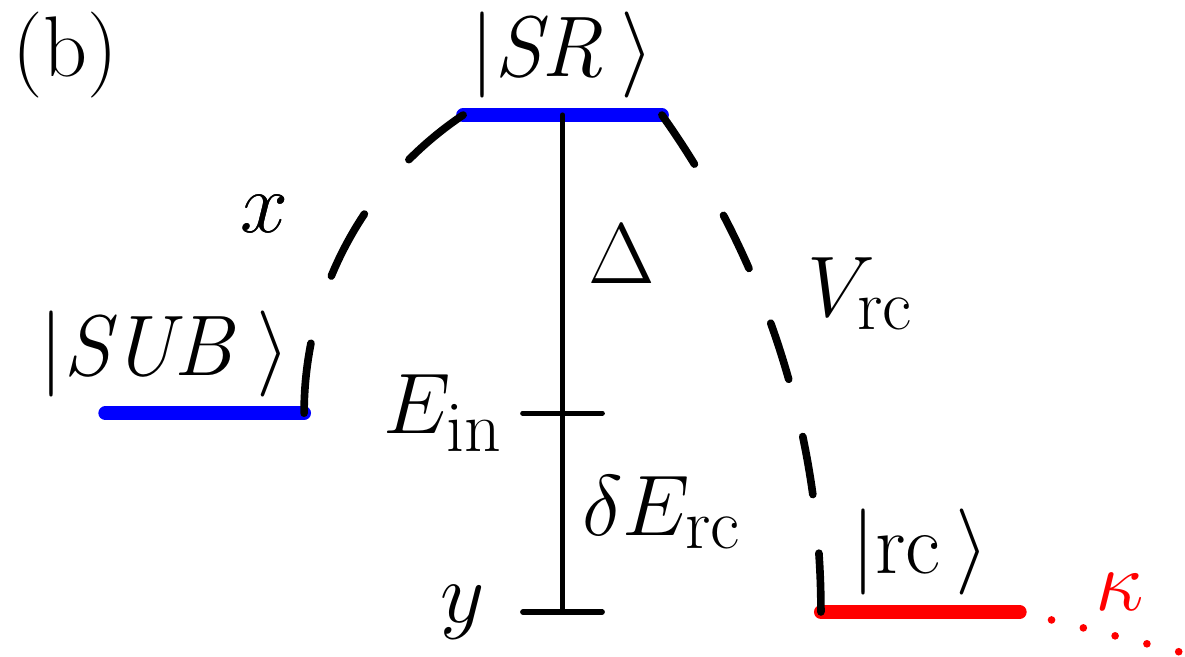}
\end{indented} \caption{In panel (a) the trimer model is shown in
  the site basis: two sites connected with the RC
with equal coupling $\Omega_{\rm rc}$ and between them with coupling
$\Omega$. In panel (b) the trimer model is shown in the
superradiant-subradiant-RC basis, see text.
}
\label{trimer}
\end{figure}

Many transport properties of systems displaying the Ring-RC structure
can be understood from the analysis of the simplest ring, i.e.\ two coupled sites, connected with a central site representing the RC, as 
depicted in figure~\ref{trimer}a.
The Hamiltonian (in site-basis) for such trimer model can be written in matrix form as follows:
\begin{equation}\label{eq:h3}
\pmatrix{
E_1 - i\Gamma_{\rm fl}/2   & \Omega &  \Omega_{\rm rc}\cr
 \Omega & E_2 - i\Gamma_{\rm fl}/2          &  \Omega_{\rm rc}         \cr
\Omega_{\rm rc}   & \Omega_{\rm rc}          &    E_{\rm rc} - i(\Gamma_{\rm fl}+\kappa)/2   
}
\end{equation}
where the action of the environment (static disorder) has been
taken into account by choosing the energy levels
$E_k$, $k=1,2$, as
Gaussian random numbers
with mean zero and variance $W^2$, and no disorder has been added to the RC site.

The loss of excitation through the RC has been described by the non-Hermitian term $-i\kappa/2$. Throughout the whole paper, $\kappa$ is assumed to be small with respect to the other coupling parameters. This choice is consistent with the realistic photosynthetic model which will be considered in the next section.
In order to make a close comparison with realistic systems and
following a standard
procedure~\cite{srfmo,srrc,prbdisorder,alberto,prbdephasing}, we also introduced 
the diagonal non-Hermitian terms $- i\Gamma_{\rm fl}/2$, with fluorescence constant $\Gamma_{\rm fl}$ much smaller than any other energy scale ($\Omega,\Omega_{\rm rc},W,\kappa$), which represent the loss of excitation from each site due to recombination.

It is convenient to move from the site-basis to the subradiant-superradiant-RC basis,
by defining the states
\[
\ket{\mathit{SUB}} =  \frac{1}{\sqrt{2}} \left( \ket{1} -\ket{2} \right)\,,
\]
\[
\ket{\mathit{SR}} = \frac{1}{\sqrt{2}} \left( \ket{1} +\ket{2} \right)\,,
\]
from which the new Hamiltonian $H$  easily follows, see also figure~\ref{trimer}b:
\begin{equation}
H=
\pmatrix{
-\Omega - i\Gamma_{\rm fl}/2  & x  &  0 \cr
& & \cr
x & \Omega - i\Gamma_{\rm fl}/2          &  \sqrt{2} \Omega_{\rm rc}       \cr
& & \cr
0   & \sqrt{2} \Omega_{\rm rc}          &    y- i(\Gamma_{\rm fl}+\kappa)/2    
}\,,
\label{eq:h3a}
\end{equation}
where the two Gaussian random variables $x=(E_1-E_2)/2$ and $y=E_{\rm rc}-(E_1+E_2)/2$ have been introduced and are such that
\[
\langle x \rangle = 0, \quad  \langle y \rangle = E_{\rm rc}, \quad \langle x^2 \rangle = 
\langle y^2 \rangle -E_{\rm rc}^2 = 
W^2/2\,.
\]
In this basis, the coupling between the subradiant state and the RC vanishes, whereas the coupling between
the superradiant state and the RC is enhanced and it is given by the
matrix element $V_{\rm rc}=\sqrt{2} \Omega_{\rm rc}$. Note that in a
ring of $N$ sites, the superradiant coupling $V_{\rm rc}$ typically
scales as $\sqrt{N}$. 
%Note that such a coupling is enhanced by a factor of $\sqrt{2}$ with respect to the coupling of each single site to the RC. This is a typical example of superradiant enhancement which typically scales as the square root of the system size.
The structure of the Hamiltonian, depicted in figure~\ref{trimer}b, implies 
that the excitation transfer from the subradiant state to the RC can only be mediated by the superradiant state, through the random coupling $x$.

Since we are interested in modeling light-harvesting complexes, a most relevant quantity 
is the efficiency, at the time $t$,  of energy transfer from the system into the RC. 
Given an initial state $\ket{\Psi_{\rm in}}$, it is defined as
\begin{equation}
\label{eq:eta}
\eta_{_{\Psi_{\rm in}}}(t)=\left\langle\kappa
\int_0^t|\bra{\rm RC}e^{-\frac{i}{\hbar}H\tau}\ket{\Psi_{\rm in}}|^2\,d\tau\right\rangle_{W}\,,  
\end{equation}
and it represents the probability of escaping out of the system  up to the time $t$. In the above definition the brackets $\langle\ldots\rangle_{W}$ indicate the average over disorder.

In numerical simulations we always consider the efficiency at a time
$t\gtrsim \hbar/\Gamma_{\rm fl}$. 
Note that the efficiency strongly depends on the time $t$ at which it is computed for $t<\hbar/\Gamma_{\rm fl}$, while it reaches a stable asymptotic value $\eta^{\infty}$ for $t\gtrsim \hbar/\Gamma_{\rm fl}$, thus motivating our choice. As $\Gamma_{\rm fl}\to 0$, the asymptotic value of the efficiency is $\eta^{\infty}=1$ for any choice of parameters. 
Hereafter we will measure energies in cm$^{-1}$ and times in ps, that corresponds to setting $1/\hbar\simeq 0.06\pi\um{cm/ps}$.

It is clear from \eqref{eq:eta} that the energy transfer efficiency is strongly dependent
 upon the initial state, a feature also studied in~\cite{fassioli}. 
Indeed, if we start from the superradiant state we are in a situation
in which there is an enhanced direct coupling to the RC at zero
disorder.  Thus, one might think
that the best situation occurs when the excitation is on the
superradiant state set at resonance with the energy of the RC. 
In this situation disorder is  only detrimental to transport,
since it tends to destroy the  superradiant
coupling~\cite{mukameldeph,prbdisorder} and it moves the system out of resonance. 
On the other side, the excitation in natural
complexes is usually spread also on subradiant states, due to the
presence of a thermal bath. Since a subradiant state is not directly
coupled to the RC, it is only through
the action of disorder that the excitation can be transferred from the initial state to the RC (i.e.\ for $W=0$, we have $\eta=0$).

We will focus our attention on this non-trivial case (see figure~\ref{trimer}b),
in which
an initial excitation is on the subradiant state at energy $E_{\rm in}=-\Omega$, coupled via $x$ to the superradiant state at energy $E_{\rm sr}=\Omega=E_{\rm in}+\Delta$, which is further coupled to the RC with tunnelling amplitude $V_{\rm rc}=\sqrt{2}\Omega_{\rm rc}$. 
Our aim is to find the system configuration which maximizes the
average transfer efficiency at $t\gtrsim \hbar/\Gamma_{\rm
  fl}$. Fixing the energy gap $\Delta=2\Omega$ between the
superradiant and the subradiant states and the superradiant-RC
coupling $V_{\rm rc}$, and assuming $\kappa$ and $\Gamma_{\rm fl}$ to be perturbative quantities, we are left with two independent parameters to be tuned to achieve the maximal efficiency: the subradiant-RC detuning $\delta E_{\rm rc}=E_{\rm rc}-E_{\rm in}$ and the strength $W$ of the random coupling $x$.

To pursue our goal we will first analyze a fully deterministic model obtained replacing the stochastic terms $x$ and $y$ in equation~\eqref{eq:h3a} with deterministic parameters as follows:
\begin{equation}
H^{\rm det}=
\pmatrix{
E_{\rm in} - i\Gamma_{\rm fl}/2  & X  &  0 \cr
& & \cr
X & E_{\rm in}+\Delta - i\Gamma_{\rm fl}/2          &  V_{\rm rc}       \cr
& & \cr
0   & V_{\rm rc}          &    E_{\rm rc} - i(\Gamma_{\rm fl}+\kappa)/2    
}\,.
\label{eq:h3d}
\end{equation}
In particular, we will discuss which pair $(\delta E_{\rm rc}^{\rm opt},X_{\rm opt})$ of values of the subradiant-RC detuning and coupling strength produce the maximal efficiency.
Subsequently, we will investigate how disorder affects those results.

In what follows, we will first analyze the results of the deterministic model and then the results in presence of disorder. In figures \ref{fig:delta0}, \ref{fig:delta1-v10}, and \ref{fig:delta20-v10} the results of the deterministic model are shown in the left panels, while those in presence of disorder, are in the right panels.

\subsection{Analysis of the deterministic model}

According to our previous assumptions, the behavior of the efficiency $\eta$ in the $(X,\delta E_{\rm rc})$-plane depends on the ratio between the two system parameters $\Delta$ and $V_{\rm rc}$. Since it is relevant
for natural systems, we 
first consider the case $\Delta=0$, which corresponds to two uncoupled ring sites in the trimer model ($\Omega=0$) equally connected to the RC. In this case, the sole energy scale of the system is $V_{\rm rc}$. We will subsequently consider the effects of a finite gap $\Delta\neq 0$.

\subsubsection{The zero-gap case}

\begin{figure}
\begin{indented}\item[]
\includegraphics[width=6.5cm]{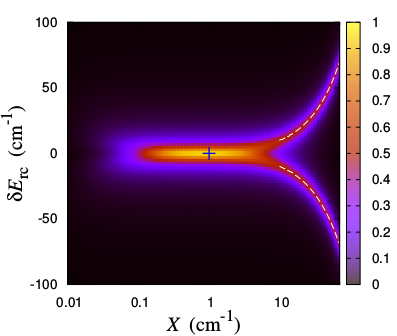}
\includegraphics[width=6.5cm]{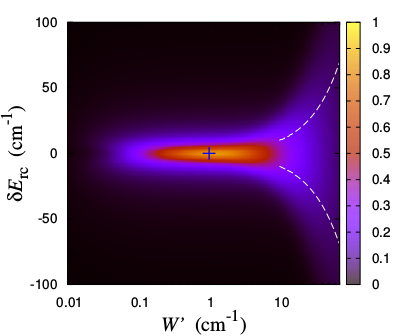}
\end{indented} \caption{Average transfer efficiency $\eta$ at fluorescence time $t_{\rm fl}=\hbar/\Gamma_{\rm fl}$, computed starting from the subradiant state $\ket{\mathit{SUB}}$, plotted in the left panel as a function of the subradiant-RC detuning $\delta E_{\rm rc}$ and of the deterministic subradiant-superradiant coupling $X$, and in the right panel as a function of $\delta E_{\rm rc}$ and of the rescaled disorder strength $W'=W/\sqrt{2}$. The blue cross indicates the estimate~\eqref{eq:gl-opt-d0} for the optimal transfer conditions. The dashed curves indicate the resonances determined in~\eqref{eq:opt-det-xlarge}. The value of the parameters are $\Delta=0\um{cm}^{-1}$, $V_{\rm rc}=1\um{cm}^{-1}$, $\kappa=0.01\um{cm}^{-1}$, and $\Gamma_{\rm fl}=10^{-4}\um{cm}^{-1}$.
}
\label{fig:delta0}
\end{figure}

With $\Delta=0$ our model corresponds to a tight-binding chain of three sites, the first (subradiant) is coupled via $X$ to the second (superradiant), with equal energy, which is coupled to the RC with strength $V_{\rm rc}$. It can be easily checked that, if we initially excite the subradiant state, the probability $P_{\rm rc}(t)$ of finding the excitation on the RC (which is a periodic function of time if we neglect the non-Hermitian terms in~\eqref{eq:h3d}) can reach the maximal value of $1$ in the shortest time when $X=V_{\rm rc}$ and $\delta E_{\rm rc}=0$, thus identifying the following global optimization condition:
\begin{equation}
\label{eq:gl-opt-d0}
X_{\rm opt}=V_{\rm rc}\qquad\text{and}\qquad \delta E_{\rm rc}^{\rm opt}=0\,.
\end{equation}
The estimate given in~\eqref{eq:gl-opt-d0}, obtained considering the
coherent transfer of excitation between the subradiant and the RC
states, is a very good estimate also of the global optimum of the
transfer efficiency, as shown in figure~\ref{fig:delta0} (blue cross
in left panel).

We observe that the condition $\delta E_{\rm rc}=0$ is not necessary for the probability of being on the RC to reach $1$, but, in combination with $X=V_{\rm rc}$, it makes such transfer the fastest. On the other hand, if we have some constraint on the coupling $X$ enforcing the condition $X\gg V_{\rm rc}$, the optimal detuning is not $\delta E_{\rm rc}=0$. 

\begin{figure}
\begin{indented}
\item[]
\includegraphics[width=6.2cm]{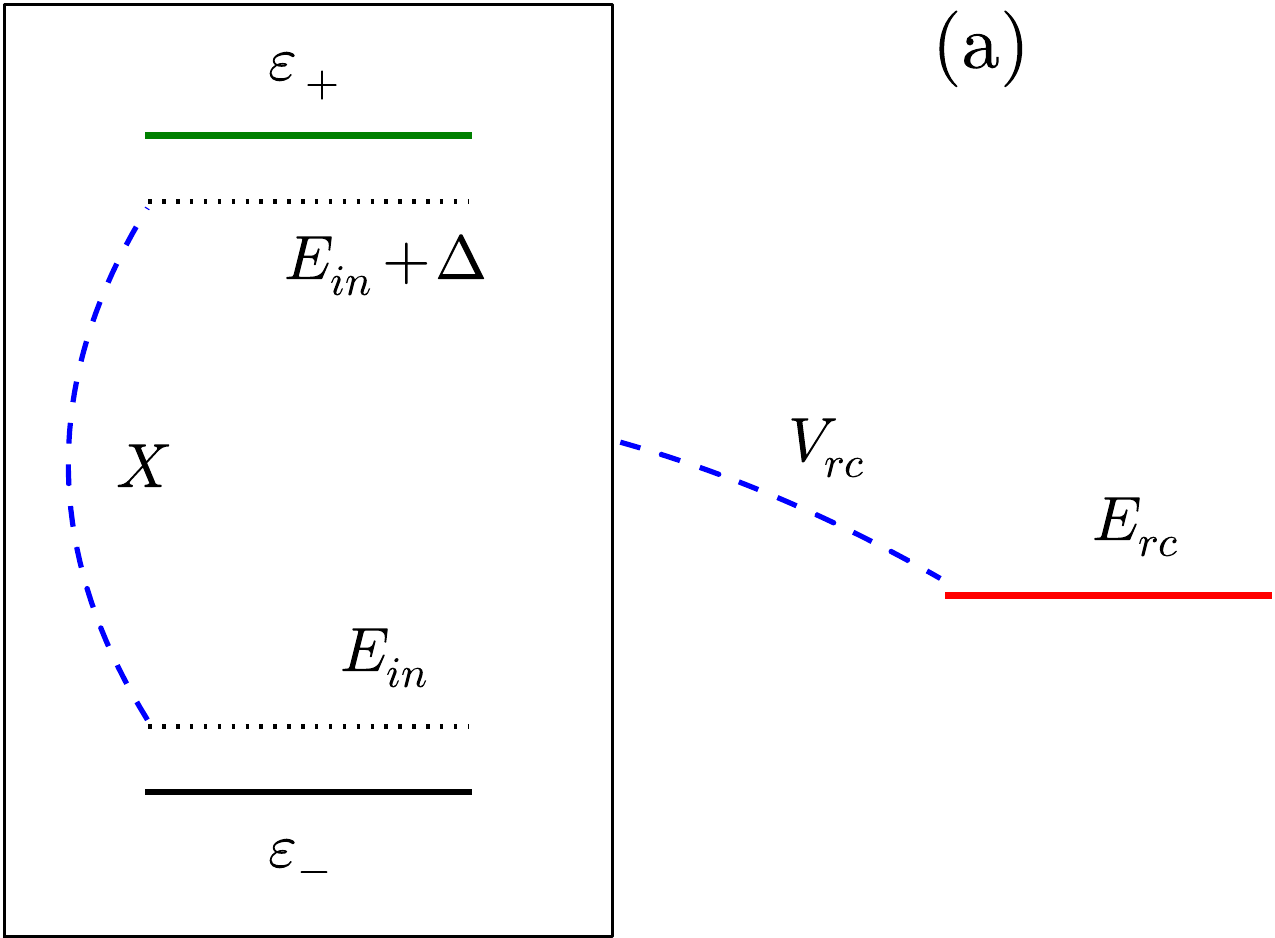}\hspace{.6cm}
\includegraphics[width=6.2cm]{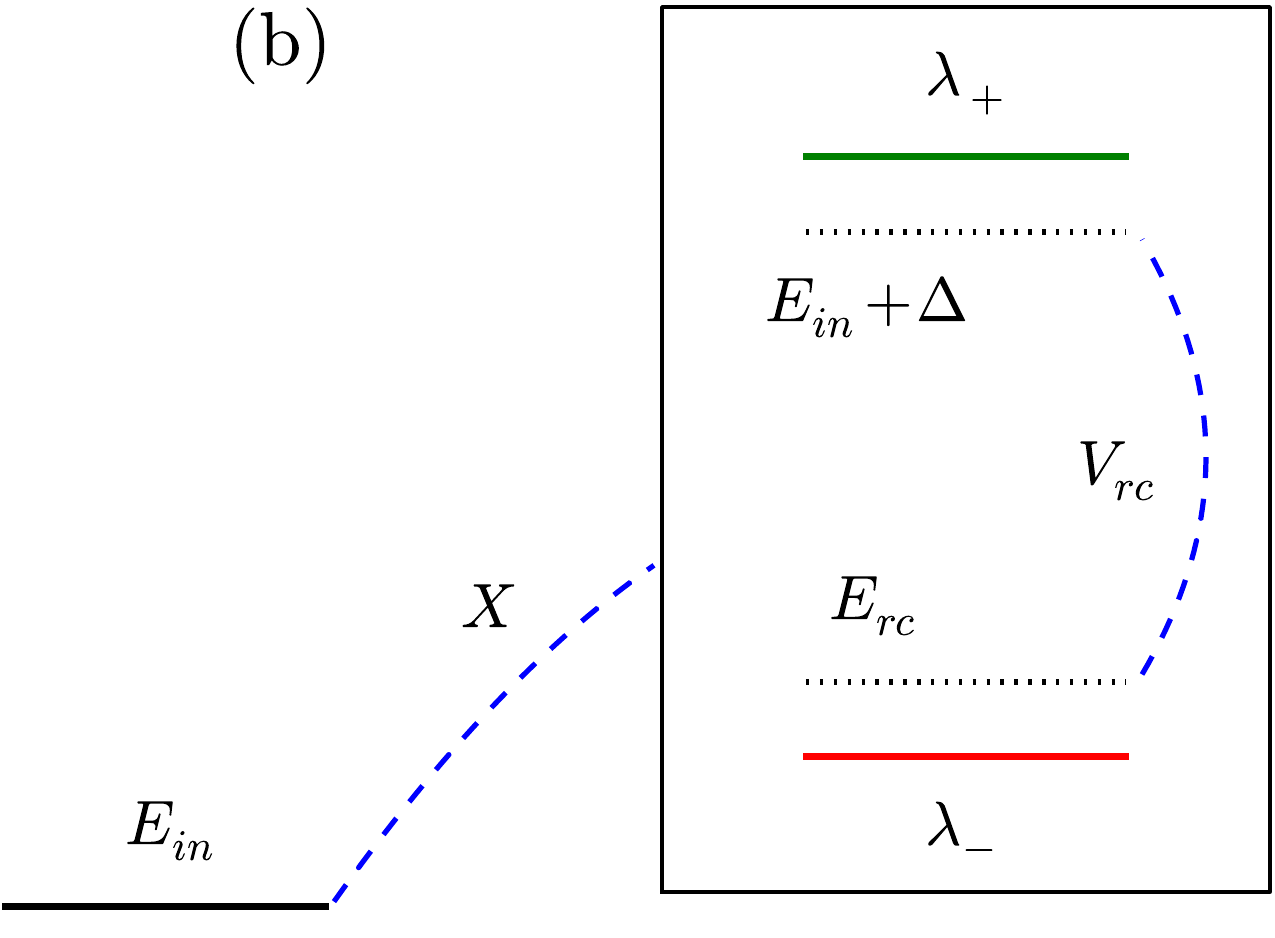}
\end{indented} \caption{Different schematic representations of the energy levels in the deterministic trimer model. 
(a) The subradiant-superradiant subsystem (framed), where the coupling $X$ produces the dressed levels with energies $\varepsilon_{\pm}$, is coupled through $V_{\rm rc}$ to the RC state at energy $E_{\rm rc}$.
(b) The initial subradiant state at energy $E_{\rm in}$ is coupled through $X$ to the superradiant-RC subsystem (framed), where the coupling $V_{\rm rc}$ produces the dressed levels with energies $\lambda_{\pm}$.}
\label{fig:2resonances}
\end{figure}

To find the resonant detuning producing the optimal transfer in the case $X\gg V_{\rm rc}$ we can consider $V_{\rm rc}$ as a perturbation, obtaining the picture illustrated in panel (a) of figure~\ref{fig:2resonances}. The subradiant and superradiant states couple and give rise to the dressed energy levels
\begin{equation}
\varepsilon_{\pm}(X)=E_{\rm in}+\frac{\Delta}{2}\pm
\sqrt{\frac{\Delta^2}{4} +X^2}\,,
\label{epsilon}
\end{equation}
which reduce to $\varepsilon_{\pm}(X)=E_{\rm in}\pm X$ for $\Delta=0$ (recall that we considered both $\kappa$ and $\Gamma_{\rm fl}$ as small perturbations, that can be neglected in finding the dressed energies). The initial excitation is equally distributed on those levels, and we can then identify two optimal detuning values by the symmetric resonant tunneling conditions
\begin{equation}
\label{eq:res-tunn-xlarge}
E_{\rm rc}=\varepsilon_{\pm}(X)\,,
\end{equation}
entailing
\begin{equation}
\label{eq:opt-det-xlarge}
\delta E_{\rm rc}(X)=\frac{\Delta}{2}\pm
\sqrt{\frac{\Delta^2}{4} +X^2}\,.
\end{equation}
For the case $\Delta=0$ we have $\delta E_{\rm rc}(X)=\pm X$ (see dashed curves in figure~\ref{fig:delta0}). Note that in~\eqref{eq:opt-det-xlarge} we kept the dependence on $\Delta$, since it will be relevant in the following.

\subsubsection{The finite-gap case}

\begin{figure}
\begin{indented}\item[]
\includegraphics[width=6.5cm]{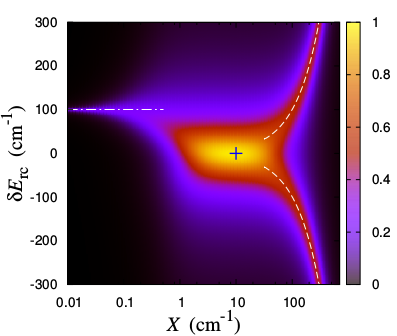}
\includegraphics[width=6.5cm]{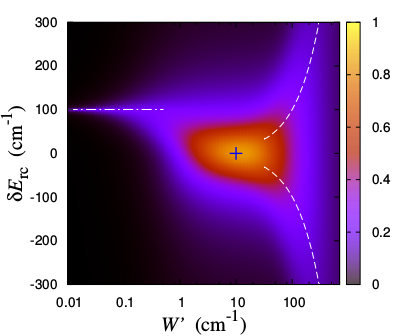}
\end{indented} \caption{Average transfer efficiency $\eta$ at fluorescence time $t_{\rm fl}=\hbar/\Gamma_{\rm fl}$, computed starting from the subradiant state $\ket{\mathit{SUB}}$, plotted in the left panel as a function of the subradiant-RC detuning $\delta E_{\rm rc}$ and of the deterministic subradiant-superradiant coupling $X$, and in the right panel as a function of $\delta E_{\rm rc}$ and of the rescaled disorder strength $W'=W/\sqrt{2}$. The blue cross indicates the estimate~\eqref{eq:gl-opt-d0} for the optimal transfer conditions. The dashed curves indicate the resonances determined in~\eqref{eq:opt-det-xlarge} and the dot-dashed line marks the the resonant condition~\eqref{eq:delta-Eres}. The value of the parameters are $\Delta=1\um{cm}^{-1}$, $V_{\rm rc}=10\um{cm}^{-1}$, $\kappa=0.01\um{cm}^{-1}$, and $\Gamma_{\rm fl}=10^{-4}\um{cm}^{-1}$.
}
\label{fig:delta1-v10}
\end{figure}

We will now investigate whether the optimal conditions given in
equation~\eqref{eq:gl-opt-d0} are valid also for finite values of the
energy gap $\Delta$. We will show that condition~\eqref{eq:gl-opt-d0}
gives a good estimate for the global optimum whenever  $V_{\rm rc}$ is
of the same order of $\Delta$ in the deterministic model, while
disorder produces some modifications, as discussed in the next subsection.

Let us first consider the situation
when the energy gap $\Delta$ is finite and $\Delta<V_{\rm rc}$. We see from figure~\ref{fig:delta1-v10} that the global optimization condition~\eqref{eq:gl-opt-d0} (blue cross) is still an excellent estimate for the configuration with maximal efficiency, and also the symmetric resonances present for $X\gg V_{\rm rc}$ follow the analytic prediction~\eqref{eq:opt-det-xlarge} (see dashed curves in figure~\ref{fig:delta1-v10}).
Nevertheless, figure~\ref{fig:delta1-v10} enlightens a somewhat unexpected feature: if we assume now the coupling $X$ to be constrained within the region $X\ll\Delta<V_{\rm rc}$, the RC energy producing the maximal efficiency, identified by a sharp resonance, is very far from either $E_{\rm in}$ or $\varepsilon_{\pm}$. 

To understand such a resonance, we can now consider $X$ as a small perturbation, exploiting the picture illustrated in panel (b) of figure~\ref{fig:2resonances}.
The superradiant and the RC states couple to give the dressed energy levels
\begin{equation}\label{eq:lambda}
\lambda_{\pm} (E_{\rm rc}) =\frac{1}{2}\left(E_{\rm in}+\Delta+E_{\rm rc}\right) \pm
\sqrt{\frac{1}{4}\left(E_{\rm in}+\Delta-E_{\rm rc}\right)^2 +V_{\rm rc}^2}\,,
\end{equation}
which clearly depend on the RC energy $E_{\rm rc}$.
The initial excitation is all on the subradiant state, since $X$ is small, and a resonant tunneling criterion would now require $E_{\rm in}$ to match the energies $\lambda_{\pm}$ of the superradiant-RC subsystem. Nevertheless, we have $E_{\rm in}<E_{\rm in}+\Delta<\lambda_+$ by construction, so that the resonant condition for $X\ll\Delta<V_{\rm rc}$ must be
\begin{equation}
\label{eq:res-tunn-xsmall}
E_{\rm in}=\lambda_-(E_{\rm rc})\,,  
\end{equation}
entailing
\begin{equation}
\label{eq:delta-Eres}
\delta E_{\rm rc}=\frac{V_{\rm rc}^2}{\Delta}\,,  
\end{equation}
which corresponds exactly to the numerical results (see dot-dashed line in figure~\ref{fig:delta1-v10}).

\begin{figure}
\begin{indented}\item[]
\includegraphics[width=6.5cm]{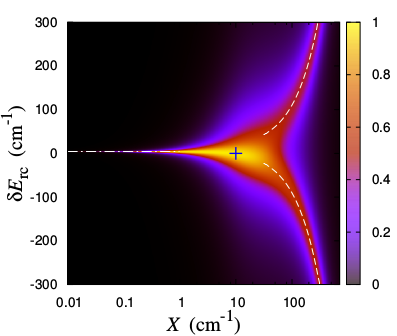}
\includegraphics[width=6.5cm]{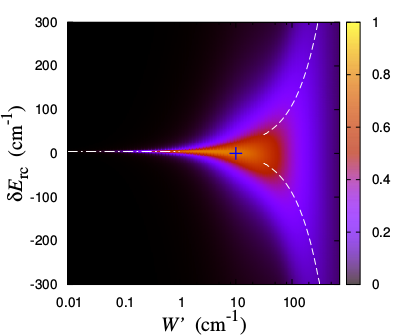}
\includegraphics[width=6.5cm]{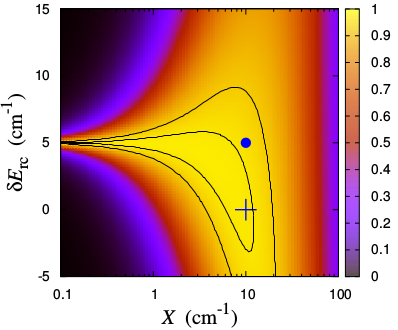}
\includegraphics[width=6.5cm]{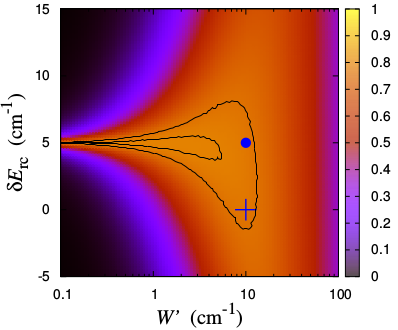}
\end{indented} \caption{Average transfer efficiency $\eta$ at fluorescence time $t_{\rm fl}=\hbar/\Gamma_{\rm fl}$, computed starting from the subradiant state $\ket{\mathit{SUB}}$, plotted in the left panels as a function of the subradiant-RC detuning $\delta E_{\rm rc}$ and of the deterministic subradiant-superradiant coupling $X$, and in the right panels as a function of $\delta E_{\rm rc}$ and of the rescaled disorder strength $W'=W/\sqrt{2}$. The blue cross indicates the estimate~\eqref{eq:gl-opt-d0} for the optimal transfer conditions. In the upper panels the dashed curves indicate the resonances determined in~\eqref{eq:opt-det-xlarge} and the dot-dashed line marks the the resonant condition~\eqref{eq:delta-Eres}. 
In the lower panels we zoomed on the high-efficiency regions to show how disorder affects the accuracy of the estimates~\eqref{eq:gl-opt-d0} and~\eqref{eq:glob-opt-disorder-l}. The blue dots mark the modification~\eqref{eq:glob-opt-disorder-s} of the estimate proposed in section~\ref{sec:dis-effect}. Isolines enclose the regions of efficiency lying within 1\% and 5\% of the maximal efficiency.
The value of the parameters are $\Delta=20\um{cm}^{-1}$, $V_{\rm rc}=10\um{cm}^{-1}$, $\kappa=0.01\um{cm}^{-1}$, and $\Gamma_{\rm fl}=10^{-4}\um{cm}^{-1}$.
}
\label{fig:delta20-v10}
\end{figure}

If we now decrease further the ratio $V_{\rm rc}/\Delta$ we find the
following remarkable result (see figure~\ref{fig:delta20-v10}, left upper
panel): the estimate~\eqref{eq:gl-opt-d0}, which was obtained for
$V_{\rm rc} \gg \Delta$, still identifies the global efficiency
optimization in the deterministic model (blue cross). Moreover, the
resonances predicted by~\eqref{eq:opt-det-xlarge} for $X\gg V_{\rm
  rc},\Delta$ and by~\eqref{eq:delta-Eres} for $X\ll V_{\rm
  rc},\Delta$ still correspond to the local optimization of the
efficiency (see white curves in figure~\ref{fig:delta20-v10}). As for
the model in presence of disorder (right upper panel of figure~\ref{fig:delta20-v10}), while the estimate~\eqref{eq:gl-opt-d0} is still within a region of significant efficiency, the optimal condition is modified by disorder, as a more detailed analysis shows (see lower panels of figure~\ref{fig:delta20-v10} and the discussion in the next subsection).

% This robustness of the estimates to the presence of the energy gap $\Delta$ is due to the fact that it only affects the frequency of the oscillations of the probability $P_{\rm rc}(t)$ of finding the excitation on the RC, without preventing it from reaching the value $1$ when $X=V_{\rm rc}$ and 
% $\delta E_{\rm rc}=0$ \textcolor{red}{The whole sentence is not clear}.

We finally observe that the width in energy $\delta E_{\rm rc}$ of the high-efficiency region at the optimal value of $X$ is proportional to the coupling $V_{\rm rc}$, providing a measure of the robustness of transfer to perturbations around the optimal detuning.

\subsection{The effect of disorder}\label{sec:dis-effect}

As can be clearly seen from figures
\ref{fig:delta0} and \ref{fig:delta1-v10}, where $V_{\rm rc}\gg\Delta$, the
results obtained from the  deterministic model are in good agreement
with those in presence of disorder.  Indeed, we can obtain an excellent estimate for the global
optimization condition by simply substituting the deterministic
coupling $X$ with $W/\sqrt{2}$ in \eqref{eq:gl-opt-d0} (blue crosses
in figures \ref{fig:delta0} and \ref{fig:delta1-v10}, right panels). 

On the other side, when $V_{\rm rc}\lesssim\Delta$ (figure
\ref{fig:delta20-v10}), disorder induces some modification of the 
global optimization condition. This effect can be clearly seen by
comparing the two lower panels of figure~\ref{fig:delta20-v10}, which
describe a situation with $V_{\rm rc}/\Delta\approx 0.5$, similar to
the one we will find in the realistic model considered below. 
By comparing the two lower panels of figure~\ref{fig:delta20-v10}, one
can see that the average over disorder shifts the optimal detuning from $\delta E_{\rm rc}=0$ to the low-disorder resonance $\delta E_{\rm rc}=V_{\rm rc}^2/\Delta$ given by~\eqref{eq:delta-Eres}.
This can be explained from the fact that the random coupling falls for
many realizations in the region $x<V_{\rm rc}\lesssim\Delta$, where
the resonance is for $\delta E_{\rm rc}=V_{\rm rc}^2/\Delta$ and not
for $\delta E_{\rm rc}=0$.  
As far as optimal disorder is concerned, even if \eqref{eq:gl-opt-d0}
overestimates its actual  value, it still gives an estimate within
$5\%$ of the maximal efficiency, see figure~\ref{fig:delta20-v10}
lower right panel. 

Note that the case $V_{\rm rc} \ll \Delta$ has not been discussed since it
is not relevant for the realistic model considered in the next section.

Summarizing,
% as long as the opening strength $\kappa$ and the fluorescence
% constant $\Gamma_{\rm fl}$ can be considered small perturbations
% with respect to the relevant system parameters (a condition
% consistent with realistic applications), 
we predict the global optimization of the average transfer efficiency $\eta(t\gtrsim \hbar/\Gamma_{\rm fl})$ from the subradiant state of the trimer into the sink placed at the RC as given by
\begin{equation}
\label{eq:glob-opt-disorder-l}
W_{\rm opt}/\sqrt{2}\simeq V_{\rm rc}\quad\text{and}\quad E_{\rm rc}=E_{\rm in}\quad\text{for}\quad V_{\rm rc}\gg\Delta\,,  
\end{equation}
and
\begin{equation}
\label{eq:glob-opt-disorder-s}
W_{\rm opt}/\sqrt{2}\simeq V_{\rm rc}\quad\text{and}\quad E_{\rm rc}=E_{\rm in}+\frac{V_{\rm rc}^2}{\Delta}\quad\text{for}\quad V_{\rm rc}\lesssim\Delta\,.  
\end{equation}
Our results also confirm the expectation that the width in energy $\delta E_{\rm rc}$ of the high-efficiency region at the optimal disorder strength is proportional to the superradiant coupling $V_{\rm rc}$.

Another physically relevant question concerns the optimal detuning
at some fixed disorder strength determined by physiological 
conditions (natural systems
are usually subject to a definite range of static disorder).
In this situation our results indicate that, 
if the disorder is constrained to be smaller than 
the energy gap $\Delta$ and the coupling $V_{\rm rc}$, 
the optimal subradiant-RC detuning is not zero, but it is always 
given by
\begin{equation}
\label{eq:delta-eres2}
\delta E_{\rm rc}=\frac{V_{\rm rc}^2}{\Delta}\,,
\end{equation}
with a significant efficiency present only in a narrow
 band around such optimal detuning.
This result is at variance with the intuitive expectation that
the best transport would be obtained 
at resonance with the initial state, $\delta E_{\rm rc}=0$.
Moreover, it can be easily
extended to $N$-sites rings  and its general validity
will be the subject of future investigations.

For large values of the disorder, as a result of the averaging procedure, we have a broad resonance centered around $\delta E_{\rm rc}=0$, which fades into two very broad resonances centered around $E_{\rm rc}=\varepsilon_\pm(W/\sqrt{2})$ given by condition~\eqref{eq:res-tunn-xlarge}, characterized by a negligible efficiency.

\section{Application to natural complexes}\label{sec:natcomp}

Let us   analyze a model of the light-harvesting complex I
(LHI) and the RC of purple bacteria.
Our purpose is to determine the optimal efficiency as a
function of both the energy detuning of the RC with respect to LHI and
the disorder strength. We will use the previous findings concerning the trimer
model to estimate such global optimum.

\subsection{LHI-RC complex of purple bacteria}

\begin{figure}
\begin{indented}\item[]
\includegraphics[width=8.5cm]{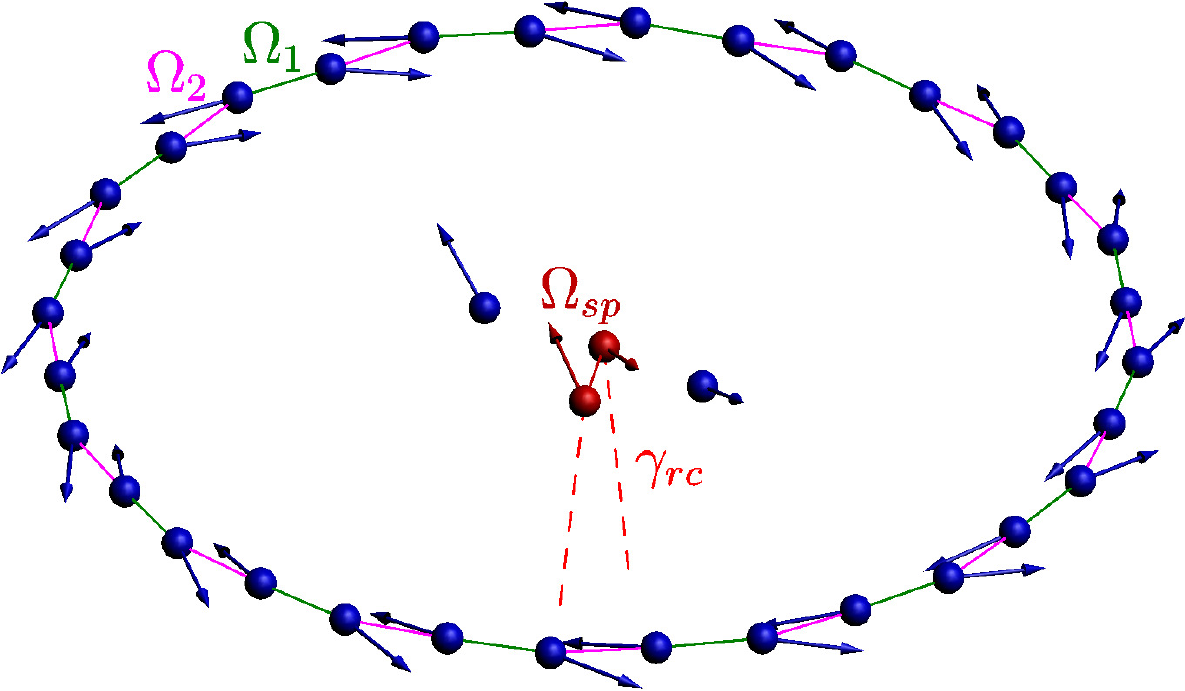}
\end{indented} \caption{Spatial structure of the LHI-RC complex. The two red sites are
the special pair ($33$, $34$) (``open'' sites) from which the excitation can escape.
The other two sites in the center are part of the RC, but they are ``closed'',
in the sense that they contribute to the interaction with the other sites
but the excitation can not escape from them.
The arrows indicate the directions of the dipole moments.}
\label{fig:dipoles}
\end{figure}

The LHI-RC complex of purple bacteria can display a variety of forms depending on the species (see~\cite{cogdell} for a review). To illustrate the relevance of our analysis in such complexes we focus on the model presented in~\cite{schultenH}. 
It consists of $32$ chromophores arranged to form the LHI ring
with equal energies $e_r$,  plus
$4$ central chromophores which represent the RC:
a special pair
with energy $e_s$  (labeled as $33,34$)
 and two accessory sites with energy $e_a$ (labeled as $35,36$), see
figure~\ref{fig:dipoles}.
The chromophores are modeled as two-level systems (sites) and, since in physiological
situations the complex is irradiated by a dilute light, we can safely apply the single-excitation approximation. The interaction
among sites is due to the transition dipole moments of the chromophores,
except for the nearest
neighbors where the dipole approximation is no longer valid. We then assume, following~\cite{schultenH}, such interaction strength to take the values $\Omega_1,\Omega_2$ alternating each site, which can be determined by fitting the energy spectrum of the LHI complex.
Moreover, the two inner sites of the RC with energy $e_s$ have also a
short-range coupling $\Omega_{\rm sp}$ determined in a similar way.

For positions $\vec{R}_s$ and  orientations  $ \vec{\mu}_s$
 of the dipole moments as well as for the nearest neighbor
 interactions ($\Omega_1,\Omega_2,\Omega_{\rm sp}$) and the site energies ($e_r, e_s, e_a$)
 we took the data (reported in Table~\ref{tab:par} and Table~\ref{tab:posdip} in the Appendix) from \cite{schultenH}.

The Hamiltonian of the system, written on the site-basis $\ket{s}$, $s=1,\ldots,36$, reads
\begin{eqnarray}
\label{eq:Hreal}
%\begin{aligned}
H_{\rm rhod}&\mbox{}=e_r\sum_{s=1}^{32}\ket{s}\bra{s}+e_{s}
\sum_{s=33}^{34}\ket{s}\bra{s}+e_{a}\sum_{s=35}^{36}\ket{s}\bra{s}\cr
&\mbox{}+\Omega_1\sum_{s=1}^{16}(\ket{2s-1}\bra{2s}+\ket{2s}\bra{2s-1})\cr
&\mbox{}+\Omega_2\sum_{s=1}^{15}(\ket{2s}\bra{2s+1}+\ket{2s+1}\bra{2s})\cr
&\mbox{}+\Omega_2(\ket{32}\bra{1}+\ket{1}\bra{32})+\Omega_{\rm sp}(\ket{33}\bra{34}+\ket{34}\bra{33})\phantom{\sum_{s=1}^{16}}\cr
&\mbox{}+C\;\widehat{\sum_{s,r}}\left[
\frac { \vec{\mu}_s  \cdot \vec{\mu}_r } {R_{sr}^3} -
\frac { 3 (\vec{R}_{sr} \cdot \vec{\mu}_s ) (\vec{R}_{sr} \cdot \vec{\mu}_r )}
{ R_{sr}^5 }\right]\,,
%\end{aligned}
\end{eqnarray}
where $\vec{R}_{sr}$
is the  vector connecting the site $s$ to  $r$,
$\vec{\mu}_s$ is the dipole moment  at the site $s$, and the constant $C$ embeds the dipole strength common to all of the chromophores.
The hat over the last summation indicates that we
consider the interaction among all sites (ring and RC) but
the terms associated with nearest neighbors along the ring and the special pair.

The loss of excitation into the RC is modeled by two independent
 sinks attached only to the sites of the special pair ($33$, $34$) with identical
 decay widths $\gamma_{\rm rc}$. Hence, the effective non-Hermitian Hamiltonian for the system is
\begin{equation}
\label{eq:Heff}
H=H_{\rm rhod}-i\frac{\gamma_{\rm rc}}{2}\sum_{s=33}^{34}\ket{s}\bra{s}\,. 
\end{equation}

The geometry of the system is illustrated in figure~\ref{fig:dipoles} and we want to
stress the fact that, while a symmetry of LHI under cyclic permutations of \emph{dimers}
can be devised, the geometry of RC sites has no symmetry whatsoever. 

\begin{figure}
\begin{indented}\item[]
\includegraphics[width=9.cm]{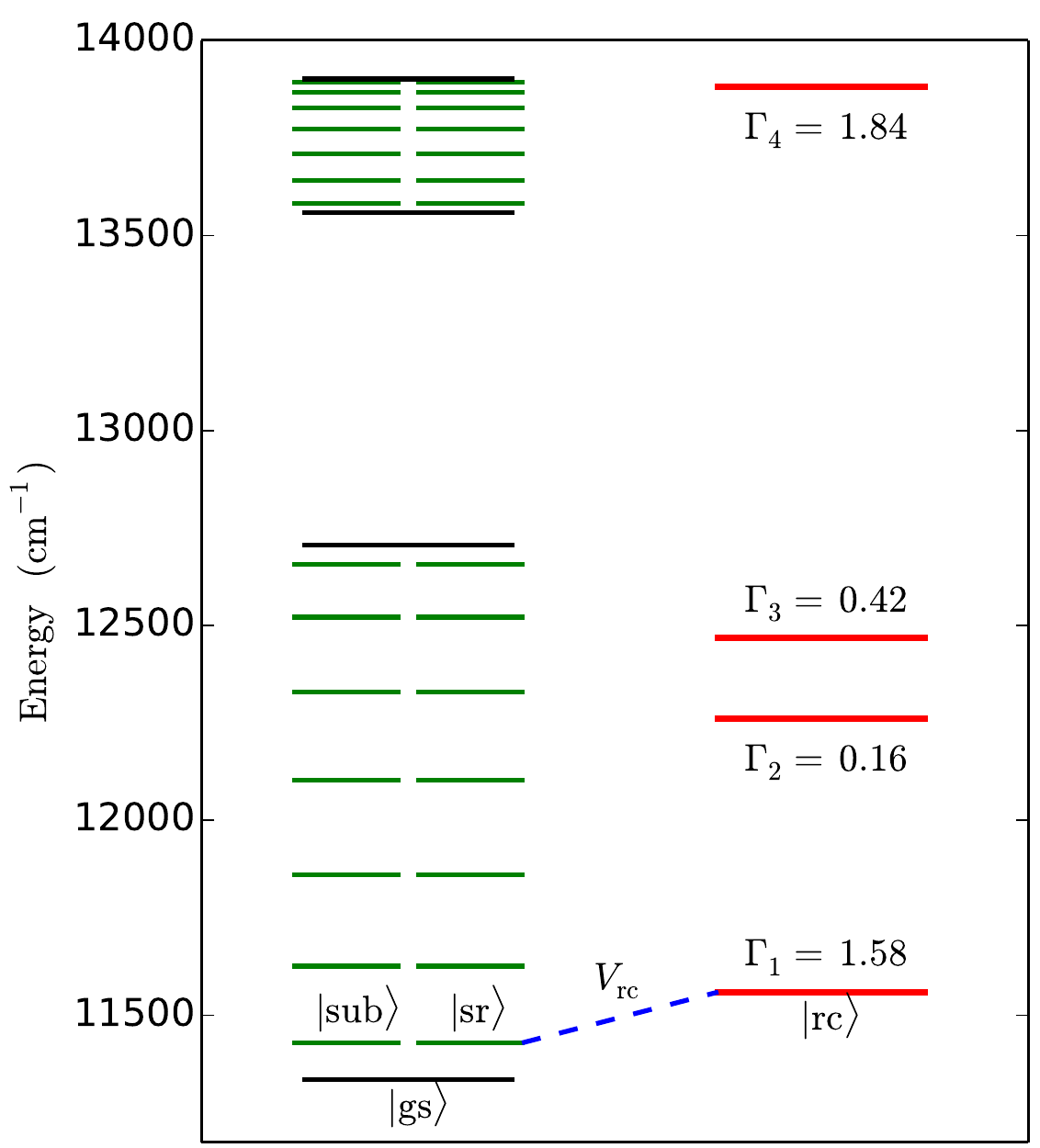}
\end{indented} \caption{Schematic representation of the 32 energy levels of the LHI complex (left) and of energies and decay widths, in cm$^{-1}$, of the four RC eigenstates (right) of purple bacteria.}
\label{fig:energylevels}
\end{figure}

Even in this realistic complex it is possible to identify LHI states that 
are superradiant/subradiant with respect to the transfer of excitation towards the RC states, 
where the excitation can be trapped. To this end, we first perform two independent diagonalizations
 of the LHI and RC sectors of the total Hamiltonian. Since the LHI sector of $H$ is Hermitian, it can be diagonalized by means of a real orthogonal matrix $R^{\rm LHI}$ (which is unitary), while the RC sector, being a non-Hermitian symmetric matrix, is diagonalized by a complex orthogonal matrix $R^{\rm RC}$ (which is non-unitary). By performing the transformation
$(R^{\rm LHI}\otimes R^{\rm RC})^t H(R^{\rm LHI}\otimes R^{\rm RC})$,
the total Hamiltonian takes the form
\[
\pmatrix{
\mathrm{diag}(E^{\rm LHI}_i) & \vec{a} & \vec{b}& \vec{c}& \vec{d}\cr
\vec{a}^t & \mathcal{E}^{\rm RC}_1& 0& 0& 0\cr
\vec{b}^t &0 & \mathcal{E}^{\rm RC}_2&0 &0 \cr
\vec{c}^t &0 &0 & \mathcal{E}^{\rm RC}_3 &0 \cr
\vec{d}^t &0 &0 &0 &\mathcal{E}^{\rm RC}_4
}
\]
where $E^{\rm LHI}_1 \leq E^{\rm LHI}_2 \leq  \ldots E^{\rm LHI}_{32}$,  are the (real) energy levels of the LHI system and
\[
\mathcal{E}^{\rm RC}_k=E^{\rm RC}_k-\frac{i}{2}\Gamma_k,\quad k=1,\ldots,4,
\]
are the complex eigenvalues of the RC system, with energy $E^{\rm RC}_k$ and decay width $\Gamma_k$.
The vectors $\vec{a}$, $\vec{b}$, $\vec{c}$, and $\vec{d}$ represent the coupling between the LHI eigenstates and the RC eigenstates. The intensity of such couplings ranges typically from $0$ to $50\um{cm}^{-1}$, with average intensity of about $6\um{cm}^{-1}$. 
The energy levels of the LHI-RC system thus obtained are illustrated in figure~\ref{fig:energylevels}.

By analyzing the coupling vectors $\vec{a}$, $\vec{b}$, $\vec{c}$, and $\vec{d}$, we can observe that, due to the peculiar symmetry structure of the system at hand, there is a great inhomogeneity in their intensity. Indeed, for any RC level we can identify a set of superradiant LHI states (strongly coupled) and a set of subradiant LHI states (weakly coupled). Such a distinction will be important in making the transport properties amenable to an analysis based on the trimer model discussed in section~\ref{sec:trimer}.

The effect of the static disorder has been 
taken into account by adding to $H$ the disorder Hamiltonian
\begin{equation}
\label{dis-re}
D=\sum_{s=1}^{36}\delta_s\ket{s}\bra{s}\,,
\end{equation}
where the $\delta_s$'s are independent Gaussian random variables centered in $0$ and with variance $W^2$.
The disorder will produce a random coupling between subradiant and superradiant states
with variance $\sim W^2/N$, where $N=32$ is the number of LHI sites.
Indeed,
the diagonalization of the LHI sector of \eqref{dis-re} 
produces a $N \times N$ full random matrix with Gaussian entries centered in $0$ and 
with variance $\sim W^2/N$, whereas the diagonalization of the RC
sector (which has $4$ sites) produces a $4\times 4$ full random matrix
with Gaussian entries centered in $0$ and  with variance $\sim W^2/4$.

It should be noted that the value $e_r$ of the LHI site energy can be nicely determined by fitting the absorption spectra of the LHI complex, whereas the RC energies $e_s$ and $e_a$ are reported to be lower than $e_r$, but their precise value is not well known and they can be shifted to match other experimental findings such as the high values of efficiency (see the discussion in~\cite{schultenH}).

The coupling with the electromagnetic field, which induces
superabsorption and superflorescence of light, has been taken into
account adding to the Hamiltonian in \eqref{eq:Heff} a
non-Hermitian part, following the procedure described in~\cite{mukameldeph}.

In the following we will keep all the parameters of the model fixed,
except for the RC energies and the strength $W$ of the disorder, with
the aim to find the global optimum. To achieve this goal, we will show
how the main features of the transport properties of the LHI-RC
complex are obtained by reducing it to two independent trimers.

\subsection{Reduction procedure}

In this  complex we have many LHI states coupled to four RC energy levels, so that it may seem very difficult to optimize the transfer efficiency by choosing the RC energies and the disorder strength independently of the LHI state initially excited.
Nevertheless, in physiological conditions the ground state and the first two degenerate excited states of the LHI
 ring are the most populated ones (together they gather 77\% 
of the total probability at $T=288\um{K}$).
Moreover, the only state of the RC we need to consider is 
the lowest energy state, since the other three RC states are
separated from it by a significant energy gap, which makes  
their effective coupling to the most populated LHI states negligible.

In the doubly degenerate subspace of the first two excited
 LHI states, if we consider the coupling to the RC, it is possible to identify a subradiant state and a superradiant one, denoted by $\ket{\rm sub}$ and $\ket{\rm sr}$, respectively (see figure~\ref{fig:energylevels}). 
Those states are such that
\[
\bra{\rm sub}H\ket{\rm rc} = a_2 = 0\,,\text{ and }
\bra{\rm sr}H\ket{\rm rc} = a_3 =-43.71\um{cm}^{-1}\,.
\]
Moreover, the LHI ground state $\ket{\rm gs}$ is subradiant with respect to transfer towards $\ket{\rm rc}$, since the coupling intensity
\[
|\bra{\rm gs}H\ket{\rm rc}| = |a_1| = 0.26\um{cm}^{-1}\,
\]
is much smaller than the average coupling, and negligible if compared to $|a_3| =43.71\um{cm}^{-1}$.

As already noted, we view the energy value $E^{\rm RC}_1$ of $\ket{\rm rc}$ as an adjustable 
parameter, which can be modified by shifting all of the RC energies
$e_s,e_a$ of the same amount. Note that the first two excited states are optically very active~\cite{schultenH}, concentrating most of the LHI fluorescence, while the
 $\ket{\rm gs}$ is optically almost inactive. On the other hand, the transfer from $\ket{\rm sr}$ towards the RC is very fast, due to
its  strong coupling. 
It is then capable to compete with (and outperform)  the decay due to fluorescence. Moreover,
when the coupling between $\ket{\rm sub}$ and $\ket{\rm sr}$ induced by disorder is strong enough, also the transfer from $\ket{\rm sub}$ becomes significant. 
For this reason, in our numerical simulations of the LHI-RC complex we took into account fluorescence in an exact way, following the approach presented in~\cite{mukameldeph}. Such approach includes the effect of superradiant loss of excitation into the electromagnetic field, thus allowing to estimate the competition between superradiant loss and superradiant transfer of the excitation. 
% For this realistic system we introduce the fluorescence
% generated by the interaction between LHI-RC sites and the electromagnetic field, which also produces the dipole interaction in $H_{\rm rhod}$.
% We then build the fluorescence (anti-Hermitian) Hamiltonian $H^{\rm fl}$ following the procedure presented in~\cite{mukameldeph}.

The problem of excitation transfer from an initial Gibbs
populations  distributed on the three lowest-energy levels towards
the lowest RC energy can be decomposed in two independent problems,
involving two trimers:
% ,
% as regards the transfer from the subradiant states $\ket{\rm gs}$ and $\ket{\rm sub}$, we consider the two trimer models associated with: 
(A) LHI ground state $\ket{\rm gs}$, superradiant $\ket{\rm sr}$, 
and $\ket{\rm rc}$; 
(B) subradiant excited state $\ket{\rm sub}$, 
superradiant $\ket{\rm sr}$, and $\ket{\rm rc}$. 

Concerning the trimer A, we neglect the direct coupling $a_1$
% \[
% \bra{\rm gs}H\ket{\rm rc} = a_1 = -0.26\um{cm}^{-1}\,
% \]
since it is much weaker than $a_3$,
and we use the model Hamiltonian~\eqref{eq:h3d} with the identifications
%can apply the results obtained in section~\ref{sec:trimer} for the weak-coupling 
%regime ($V_{\rm rc} \lesssim \Delta$),
%since
\[
\Delta=E^{\rm LHI}_3-E^{\rm LHI}_1= 94 \ {\rm cm }^{-1}\,\text{ and }
% \]
% and
% \[
V_{\rm rc}=|a_3| =43.71\um{cm}^{-1}\,.
\]
As for trimer B, we can use again the model Hamiltonian~\eqref{eq:h3d} with $\Delta\simeq 0$, since
\[
E^{\rm LHI}_2=E^{\rm LHI}_3=11429\um{cm}^{-1}\,,
\]
and $V_{\rm rc}=|a_3|$. In both models the opening strength $\kappa$ in~\eqref{eq:h3d} is given by $\Gamma_1=1.58\um{cm}^{-1}$.
Moreover, 
the variance of the random coupling between subradiant states and $\ket{\rm sr}$, which 
will be important in what follows, is of the order of $W^2/N$ (with $N=32$).

% numerically found to be 
% $\sigma_{\rm sub}^2 = c_1 W^2$,  ($c_1\simeq 0.05$), while we set the variance 
% of the random coupling between $\ket{\rm gs}$ and $\ket{\rm sr}$ 
%  to be $\sigma_{\rm gs}^2=W^2/N$ where $N=32$ is the number of sites.

\subsection{Optimality conditions}

At room temperature the excitation is evenly shared among the states
$\ket{\rm gs}$ (33\%), $\ket{\rm sub}$ (22\%) and $\ket{\rm sr}$
(22\%). To find the optimal conditions we will focus on the
subradiant states, which are the only ones to display a non-monotone
behavior with disorder. Moreover, we can use the conditions found in 
section \ref{sec:trimer} for the trimer model.

\subsubsection*{Transfer from $\ket{\rm gs}$.}
%We observe that $E^{\rm RC}_1$ is often taken equal to $E^{\rm LHI}_1$ (see~\cite{schultenH}) by invoking a resonant tunneling principle associated only with the direct coupling
% between $\ket{\rm gs}$ and $\ket{\rm rc}$. 
%This is done to achieve a maximal energy transfer efficiency, but we
%will show that the coupling between $\ket{\rm gs}$ and the
%superradiant state $\ket{\rm sr}$ induced by disorder cannot be
%neglected. Indeed, 
Given that $|a_1|\ll |a_3|$, the indirect connection between $\ket{\rm gs}$ and $\ket{\rm rc}$ through $\ket{\rm sr}$ is the most relevant. We then expect that, to find the optimal value of $E^{\rm RC}_1$ relative to transfer from $\ket{\rm gs}$, one must consider the trimer A, which predicts the optimal resonant configuration at 
\[
E^{\rm RC}_1=E_{\rm res}^{\rm gs}=E^{\rm LHI}_1+\frac{|a_3|^2}{E^{\rm LHI}_3-E^{\rm LHI}_1}\approx 11355\um{cm}^{-1} \,,
\]
since $V_{\rm rc}/\Delta\approx 0.5$ (see conditions~\eqref{eq:glob-opt-disorder-s}).
Such a configuration can be produced by setting the energy of the special pair
$e_s = E^{\rm RC}_1 +1188\um{cm}^{-1}=12543\um{cm}^{-1}$.
The robustness with respect to detuning of such optimum has been
discussed in section~\ref{sec:trimer}, indicating that $E^{\rm RC}_1$
should lie within an interval $\pm V_{\rm rc}$ around $E_{\rm
  res}^{\rm gs}$. As explained in section \ref{sec:trimer}, the
optimal disorder is given by the condition $\sigma_{\rm gs}\simeq
W/\sqrt{N} \simeq V_{\rm rc}=|a_3|$, which produces the estimate $W_{\rm opt}\simeq 250\um{cm}^{-1}$.
The optimal transfer conditions for trimer A are then:
\[
W\simeq 250\um{cm}^{-1}\quad\text{and}\quad E^{\rm RC}_1=E_{\rm res}^{\rm gs}\pm V_{\rm rc}\simeq(11355\pm 44)\um{cm}^{-1}\,.
\]

\subsubsection*{Transfer from $\ket{\rm sub}$.}
In this case ($\Delta=0$) the optimal detuning is zero (see
conditions~\eqref{eq:glob-opt-disorder-l}), entailing $E^{\rm
  RC}_1=E^{\rm LHI}_3\pm V_{\rm rc}$, and the optimal disorder is
again given by $\sigma_{\rm sub}\simeq W/\sqrt{N} \simeq V_{\rm rc}$. So that the optimal conditions for this case are
\[
W\simeq 250\um{cm}^{-1}\quad\text{and}\quad E^{\rm RC}_1=E_{\rm res}^{\rm gs}\pm V_{\rm rc}\simeq(11429\pm 44)\um{cm}^{-1}\,.
\]

%392\pm 7

\subsubsection*{Global conditions.}

From the discussion above we see that the global optimization conditions are somewhat different for each of the two LHI states we analyzed, but the cooperatively enhanced superradiant coupling $V_{\rm rc}$ is large enough to allow for a significant overlap of the high-efficiency regions pertaining to each initial state. This overlap identifies the interval
\begin{equation}
E^{\rm RC}_1\simeq (11392\pm 7)\um{cm}^{-1}\,,
\end{equation}
corresponding to $e_s = (12580\pm 7)\um{cm}^{-1}$.
On the other side, from the above discussion it is clear that the optimal disorder range is $W\simeq 250\um{cm}^{-1}$.

% We then estimate that the optimal disorder lies between the $100\um{cm}^{-1}$ at which superradiance is hindered and the $200$--$250\um{cm}^{-1}$ that maximize transport from the subradiant states. Similarly, we expect to find a simultaneous optimization by placing $E^{\rm RC}_1$ half-way between $E^{\rm LHI}_3$ (optimal for $\ket{\rm sub}$ and $\ket{\rm sr}$) and $E_{\rm res}^{\rm gs}$ (optimal for $\ket{\rm gs}$). These conditions correspond to the estimates
% \begin{equation}
% E^{\rm RC}_1\simeq 11392\um{cm}^{-1}\,,
% \end{equation}
% corresponding to $e_s = 12580\um{cm}^{-1}$, and
% \begin{equation}
% W_{\rm opt}\simeq 150\um{cm}^{-1}\,.
% \end{equation}

% should be less than $100\um{cm}^{-1}$ for the first two cases, while larger than $90\um{cm}^{-1}$ for the last case, so that we can estimate the global optimal disorder to be approximately $100\um{cm}^{-1}$. For such a choice of the optimal disorder, the optimal detuning for the ground state should lie between $10$ and $30$ above $E^{\rm LHI}_1$ (a range centered around $E_{\rm res}^{\rm gs}$). The upper bound of this interval lies close to the lower bound of the optimal detuning for the other two cases, so that our estimate for the global optimal conditions are
% \begin{equation}
% E^{\rm RC}_1\simeq E_{\rm res}^{\rm gs}+\frac{\sigma^{\rm opt}_{\rm gs}|a_3|}{E^{\rm LHI}_3-E^{\rm LHI}_1}\simeq 11365\um{cm}^{-1}\,,
% \end{equation}
% corresponding to $e_s = 12553\um{cm}^{-1}$, and
% \begin{equation}
% W_{\rm opt}\simeq 100\um{cm}^{-1}\,.
% \end{equation}

\begin{figure}
\begin{indented}\item[]
\includegraphics[width=11.cm]{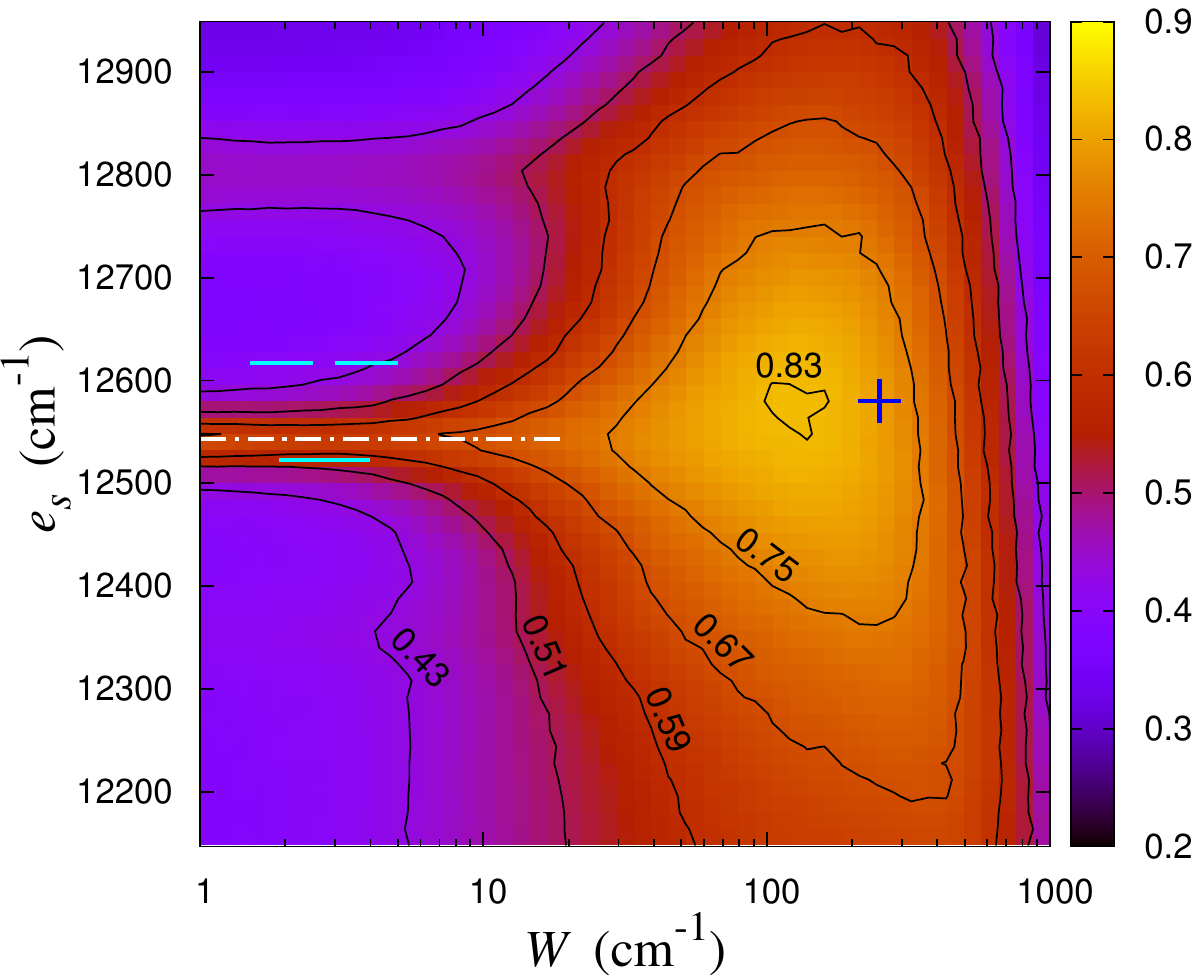}
\end{indented} \caption{Average transfer efficiency at $20\um{ns}$ for
  the LHI-RC complex, starting from a Gibbs distribution of LHI
  eigenstates at temperature $T=288\um{K}$, plotted as a function of
  the disorder strength $W$ and of the energy $e_s$ of the special
  pair. The horizontal dot-dashed line indicates the value of
  $e_s=12543\um{cm}^{-1}$ for which $E^{\rm RC}_1=E_{\rm res}^{\rm
    gs}$ (resonant configuration). The blue cross marks the estimated
  optimal conditions, see text. The values of $e_s$ such that $E^{\rm RC}_1=E^{\rm LHI}_j$, $j=1,2,3$, (possible resonances with the first three LHI eigenstates) are marked with cyan bars.}
\label{fig:maxreal}
\end{figure}

This estimate of the global optimal conditions is shown as a blue
cross in figure~\ref{fig:maxreal} and it is very close to the
region of maximal efficiency found numerically. Note that, while the detuning energy is in
excellent agreement with the numerical results, the optimal disorder
slightly overestimates the actual maximal value. This can be due to
the fact that the efficiency of the superradiant state (not considered here) decreases with
disorder, thus producing a lowering of the optimal disorder
value. Another possible cause of deviation might be that for the
ground state the optimal disorder can be lower than what we estimated,
as emerges from the discussion about figure~\ref{fig:delta20-v10}, lower right panel.
Note that the optimal disorder strength falls within the range 
of the natural disorder strengths ($30\um{cm}^{-1}$ to $300\um{cm}^{-1}$~\cite{fassioli,vangrondelle}).
% \textcolor{red}{ Io qui cambierei un po' figura e descrizione. Mi spiego:
% abbiamo ottenuto 2 valori ottimali di W e detuning, fare la media
% non mi sembra abbia molto senso (perchè dovrebbero avere lo stesso peso?)
% Io invece rinuncerei a mettere una croce ottimale e vedrei meglio
% mettere una regione ottimale dashed white rectangle di noise
% e detuning in base a quanto vistot in precedenza e farei notare 
% che si overlappa perfettamente con la regione di efficienza ottimale.
% Lascerei inoltre perdere la visualizzazione grafica della regione
% naturale dei noise, possiamo dire che i noise ottimali trovati giacciono
% proprio nella regione naturale che e'...... Questo mi sembra piu'
% onesto e sensato che dividere per due. Sempre nella figura lascerei
% perdere la linea di risonanza orizzontale mentre lascerei i 3 livelli.
% Il testo andrebbe moi modificato di conseguenza insieme alla
% figura caption.} 

The results presented in figure~\ref{fig:maxreal} also show that, for small disorder strength, the transfer of excitation from the LHI ring into the RC has an optimal efficiency when $E^{\rm RC}_1=E_{\rm res}^{\rm gs}$ (horizontal dot-dashed line), consistent with our analysis,
whereas both the configurations with $E^{\rm RC}_1=E^{\rm LHI}_1$ or $E^{\rm RC}_1=E^{\rm LHI}_2=E^{\rm LHI}_3$ (marked with cyan bars in figure~\ref{fig:maxreal}) would produce a lower efficiency for small $W$. 
%This fact proves that, to optimize transfer, it is important to consider our resonant tunneling condition for an effective trimer model, rather than the standard resonance argument, that would work only for two directly coupled states.
%Note that, having introduced fluorescence according to~\cite{mukameldeph}, we obtain an average fluorescence time for the LHI structure of 
%about $14\um{ns}$; hence, we computed the transfer efficiency at the slightly larger time of $20\um{ns}$, given 
%by the evolution of LHI states under the total Hamiltonian $H+D+H^{\rm fl}$.
Note that we computed the transfer efficiency at $t=20\um{ns}$, a time larger than the average fluorescence time.

\section{Conclusions}\label{sec:conclusions}

We considered both paradigmatic and realistic models of quantum networks in which a ring of 
peripheral sites is coupled to some inner sites, where the excitation can be 
trapped in a reaction center. In similar networks the ring states can often be classified 
as superradiant or subradiant, based on how they transfer the excitation into the reaction center.

Subradiant states are not directly coupled to the reaction center, but static on-site
disorder can effectively couple them 
with superradiant states. This opens an indirect path for the transfer of excitation
from subradiant states to the reaction center mediated by the superradiant state. 
The static disorder which activates the transfer from subradiant states, when too strong,
hinders transport so that an optimal disorder condition can be determined.

We identify the building block of such kind of transport: a trimer
with a subradiant initial state, a target RC state, and a superradiant
state coupled to the RC and, via static disorder, to the subradiant state. Four parameters determine the different regimes in which the trimer can operate: the energy distance $\Delta$ between the subradiant and the superradiant state, the intensity $W$ of the random coupling between those states, the direct coupling $V_{\rm rc}$ between the superradiant state and the RC, and the detuning $\delta E_{\rm rc}$ between the subradiant initial state and the RC.

We study how to optimize the energy transfer efficiency by varying both the disorder strength $W$ and the detuning $\delta E_{\rm rc}$. We determine such optimal conditions as given by: (i) $W\simeq V_{\rm rc}$ and $\delta E_{\rm rc}=0$ for $V_{\rm rc}\gg\Delta$; (ii) $W\simeq V_{\rm rc}$ and $\delta E_{\rm rc}={V_{\rm rc}^2}/{\Delta}$ for $V_{\rm rc}\lesssim\Delta$. Note that the optimal disorder is determined by the superradiant coupling $V_{\rm rc}$.
We also discuss how to determine the optimal detuning when 
the disorder strength is constrained to be much smaller 
or much larger than both $\Delta$ and $V_{\rm rc}$.

Notably, our general results can be applied to a realistic model for the light-harvesting 
complex (LHI) and reaction center (RC) of purple bacteria.
A reduction of the whole systems to a set 
of independent trimers allows for a
detailed explanation of the whole process of transfer from an initial Gibbs thermal state
and  justifies the  very high 
efficiency of excitation transfer even at room temperature. Our analysis points towards a 
possible correlation between the structure of such a system and transfer optimization. 
Indeed, the superradiant coupling generated by the symmetry of the LHI complex is strong enough to allow for a simultaneous optimization of transfer from the relevant LHI eigenstates. Moreover, it entails an optimal disorder
 strength for the transfer from such Gibbs distribution of initial states 
that falls within the estimated range of physiological disorder. 
We believe that
our analysis covers  those quantum networks in which a subradiant
subspace
 (also known in literature as trapping-free subspace~\cite{invsub}) 
is present. Especially to understand the enhancement in transport 
due to disorder, which is ubiquitous in such structures, it is necessary to go beyond the standard perturbative analysis.

It will be important to consider, in a future work, the effect of different baths, represented by 
noise with short correlation time, which induce loss of coherence during the 
quantum evolution. Indeed, the different nature of such kind of noise, always present 
in physiological situations, and their interplay with static disorder may entail different estimates for the optimal noise strength, 
while we expect the resonant tunneling criterion to remain as presented in this paper.

\ack  %command only for IOP
This work was partially supported by the research promotion initiative of Universit\`a Cattolica del Sacro Cuore.
We also acknowledge useful discussions with Lev Kaplan, Debora Contreras-Pulido, and
Robin Kaiser.

\appendix

\section{Data for the realistic model}

The values of the parameters entering the definition of Hamiltonian~\eqref{eq:Hreal} for the realistic model of the LHI-RC complex of the purple bacterium \emph{Rhodobacter Sphaeroides} have been taken from~\cite{schultenH} and are reported in Tables~\ref{tab:par} and \ref{tab:posdip}.

\begin{table}
\caption{Energy and coupling parameters of the model for the LHI-RC complex.}
\begin{indented}
\lineup
\item[]\begin{tabular}{@{}lll}
\br
Description & Symbol & Value \\
\mr
energy of the ring sites & $e_r$ & \012911 $\mathrm{cm}^{-1}$\\
energy of the special pair ($33$, $34$) & $e_s$ & \012748 $\mathrm{cm}^{-1}$\\
energy of the accessory pair ($35$, $36$) & $e_a$ & \012338 $\mathrm{cm}^{-1}$\\
coupling within a dimer $(2s+1,2s)$ & $\Omega_1$ & \0\0\0806 $\mathrm{cm}^{-1}$\\
coupling across two dimers $(2s,2s+1)$ & $\Omega_2$ & \0\0\0377 $\mathrm{cm}^{-1}$\\
coupling within the special pair ($33$, $34$) & $\Omega_{\rm sp}$ & \0\01000 $\mathrm{cm}^{-1}$\\
dipole strength of all of the chromophores & $C$ & 519310 $\text{\AA}^3\mathrm{cm}^{-1}$\\
\br
\end{tabular}
\end{indented}
\label{tab:par}
\end{table}

\begin{table}
\caption{Position and transition-dipole direction of the LHI-RC sites.}
\begin{indented}
\lineup
\item[]\begin{tabular}{@{}lllllll}
\br
Site & $x\text{ (\AA)}$ & $y\text{ (\AA)}$ & $z\text{ (\AA)}$ & $\mu_x$ & $\mu_y$ & $\mu_z$ \\
\mr
\01& 44.706 &\-12.591&71.986 & 0.634& 0.760   &0.147 \\
\02& 47.167&      \0\-3.677& 72.184 & \-0.452&\-0.886  &0.098 \\
\03& 46.122&      \05.475&  71.986 & 0.295& 0.944   &0.147 \\
\04& 44.984&      14.653& 72.184 & \-0.079&\-0.992  &0.098 \\
\05& 40.515&      22.709& 71.986 & \-0.089&0.985   &0.147 \\
\06& 35.952&      30.752& 72.184 & 0.307& \-0.947  &0.098 \\
\07& 28.741&      36.485& 71.986 & \-0.459&0.876   &0.147 \\
\08& 21.448&      42.169& 72.184 & 0.646& \-0.757  &0.098 \\
\09& 12.591&      44.706& 71.986 & \-0.760&0.634   &0.147 \\
10& \03.677&      47.167& 72.184 & 0.886& \-0.452  &0.098 \\
11& \0\-5.475&     46.122& 71.986 &\-0.944& 0.295   &0.147  \\
12& \-14.653&44.984&     72.184 & 0.992& \-0.079  &0.098 \\
13& \-22.709     &40.515&71.986 & \-0.985&\-0.089  &0.147 \\
14& \-30.752&35.952&     72.184 & 0.947& 0.307   &0.098 \\
15& \-36.485&28.741&     71.986 & \-0.876&\-0.459  &0.147 \\
16& \-42.169&21.448&     72.184 & 0.757& 0.646   &0.098 \\
17& \-44.706&12.591&     71.986 & \-0.634&\-0.760  &0.147 \\
18& \-47.167&\03.677&      72.184 & 0.452& 0.886   &0.098 \\
19& \-46.122&\0\-5.475&     71.986 & \-0.295&\-0.944  &0.147 \\
20& \-44.984&\-14.653&72.184 & 0.079&     0.992   &0.098 \\
21& \-40.515&\-22.709     &71.986 & 0.089&\-0.985  &0.147 \\
22& \-35.952&\-30.752&72.184 & \-0.307&0.947       &0.098 \\
23& \-28.741&\-36.485&71.986 &0.459&      \-0.876  &0.147  \\
24& \-21.448&\-42.169&72.184 & \-0.646&0.757       &0.098 \\
25& \-12.591&\-44.706&71.986 & 0.760&     \-0.634  &0.147 \\
26& \0\-3.677&     \-47.167&72.184 &\-0.886& 0.452   &0.098  \\
27& \05.475&      \-46.122&71.986 & 0.944& \-0.295  &0.147 \\
28& 14.653&     \-44.984&72.184 & \-0.992&0.079   &0.098 \\
29& 22.709&     \-40.515&71.986 & 0.985& 0.089   &0.147 \\
30& 30.752&     \-35.952&72.184 & \-0.947&\-0.307  &0.098 \\
31& 36.485&     \-28.741&71.986 & 0.876& 0.459   &0.147 \\
32& 42.169&     \-21.448&72.184 & \-0.757&\-0.646  &0.098 \\
\mr
33& \0\-3.402&     \0\-2.049& 68.512 & \-0.379&\-0.860  &\-0.342 \\
34& \03.317&      \02.301&  68.262 & 0.371& 0.871   &\-0.321 \\
\mr
35& \03.369&      \0\-9.909& 72.059 & 0.208& 0.854   &\-0.476 \\
36& \0\-2.852&     10.539& 70.886 & \-0.300&\-0.870  &\-0.391 \\
\br
\end{tabular}
\end{indented}
\label{tab:posdip}
\end{table}

\section*{References} %command only for IOP

\end{document}